\documentclass{article}

\usepackage{arxiv}

\usepackage[utf8]{inputenc} 
\usepackage[T1]{fontenc}    

\usepackage{url}            
\usepackage{booktabs}       
\usepackage{nicefrac}       
\usepackage{amsmath,amsfonts}
\usepackage{microtype}      
\usepackage{hyperref}       
\usepackage{cleveref}       
\usepackage{lipsum}         
\usepackage{natbib}
\usepackage{doi}

\usepackage{algpseudocode}
\usepackage{algorithm}
\usepackage{array}
\usepackage[caption=false,font=normalsize,labelfont=sf,textfont=sf]{subfig}
\usepackage{multirow}
\usepackage{siunitx}
\usepackage{textcomp}
\usepackage{stfloats}
\usepackage{url}
\usepackage{verbatim}
\usepackage{booktabs} 
\usepackage{graphicx}
\usepackage{cite}

\title{SeqTG: Scalable Combinatorial Test Generation via Sequential Integer Linear Programming}

\newif\ifuniqueAffiliation
\uniqueAffiliationtrue

\ifuniqueAffiliation 
\author{ \href{https://orcid.org/0000-0000-0000-0000}{\includegraphics[scale=0.06]{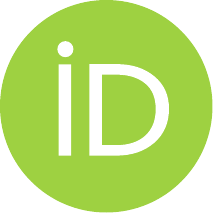}\hspace{1mm}Sitong Yang} \\
	School of Computer Science and Artificial Intelligence\\
	Hubei University of Technology\\
	Wuhan 430068, China \\
	\texttt{y18071971380@126.com} \\
	\And
	\href{https://orcid.org/0000-0000-0000-0000}{\includegraphics[scale=0.06]{orcid.pdf}\hspace{1mm}Wanying Bao} \\
	School of Computer Science and Artificial Intelligence\\
	Hubei University of Technology\\
	Wuhan 430068, China \\
	\texttt{22103011220@hbut.edu.cn} \\
	\And
	\href{https://orcid.org/0000-0000-0000-0000}{\includegraphics[scale=0.06]{orcid.pdf}\hspace{1mm}Yinyin Song} \\
	School of Computer Science and Artificial Intelligence\\
	Hubei University of Technology\\
	Wuhan 430068, China \\
	\texttt{n3476220435@126.com} \\
	\And
	\href{https://orcid.org/0000-0000-0000-0000}{\includegraphics[scale=0.06]{orcid.pdf}\hspace{1mm}Yueting Cheng} \\
	School of Computer Science and Artificial Intelligence\\
	Hubei University of Technology\\
	Wuhan 430068, China \\
	\texttt{2310300632@hbut.edu.cn} \\
	\And
	\href{https://orcid.org/0000-0000-0000-0000}{\includegraphics[scale=0.06]{orcid.pdf}\hspace{1mm}Qian Li} \\
	School of Computer Science and Artificial Intelligence\\
	Hubei University of Technology\\
	Wuhan 430068, China \\
	\texttt{2310300635@hbut.edu.cn} \\
	\And
	\href{https://orcid.org/0000-0000-0000-0000}{\includegraphics[scale=0.06]{orcid.pdf}\hspace{1mm}Chao Wei}\thanks{Chao Wei is the first Corresponding author.} \\
	School of Computer Science and Artificial Intelligence\\
	Hubei University of Technology\\
	Wuhan 430068, China \\
	\texttt{weichao.2022@hbut.edu.cn} \\
}
\else
\usepackage{authblk}

\newbox{\orcid}\sbox{\orcid}{\includegraphics[scale=0.06]{orcid.pdf}} 
\author[1]{%
	\href{https://orcid.org/0000-0000-0000-0000}{\usebox{\orcid}\hspace{1mm}David S.~Hippocampus\thanks{\texttt{hippo@cs.cranberry-lemon.edu}}}%
}
\author[1,2]{%
	\href{https://orcid.org/0000-0000-0000-0000}{\usebox{\orcid}\hspace{1mm}Elias D.~Striatum\thanks{\texttt{stariate@ee.mount-sheikh.edu}}}%
}
\affil[1]{Department of Computer Science, Cranberry-Lemon University, Pittsburgh, PA 15213}
\affil[2]{Department of Electrical Engineering, Mount-Sheikh University, Santa Narimana, Levand}
\fi

\renewcommand{\shorttitle}{\textit{arXiv} Template}

\hypersetup{
	pdftitle={A template for the arxiv style},
	pdfsubject={q-bio.NC, q-bio.QM},
	pdfauthor={David S.~Hippocampus, Elias D.~Striatum},
	pdfkeywords={First keyword, Second keyword, More},
}

\begin{document}
	\maketitle
	
	\begin{abstract}
		Combinatorial Testing (CT) is essential for detecting interaction-triggered faults, yet generating minimal Covering Arrays under complex constraints remains an unresolved NP-hard challenge. Current greedy algorithms are highly scalable but suffer from severe ``diminishing returns'': they efficiently cover initial interactions but produce bloated, redundant test suites when struggling to pack the final few difficult pairs. While exact mathematical programming could theoretically address this inefficiency, it has historically been intractable due to combinatorial explosion. In this paper, we pioneer the application of exact mathematical modeling to CT by introducing SeqTG, a scalable framework based on Sequential Integer Linear Programming (ILP). To circumvent the scalability barrier, SeqTG employs a novel Warm-Start strategy: a rapid greedy initialization first clears the ``easy'' interactions, allowing the rigorous ILP solver to exclusively optimize the fragmented, difficult-to-cover remainder. The pipeline operates in three stages: (1) a Constraint-First phase grouping must-include requirements via graph partitioning; (2) an Incremental Optimization phase targeting the remaining interactions with sequential ILP; and (3) a Global Minimization phase eliminating redundancies via set-covering. Extensive evaluations across standard benchmarks and 200 large-scale configurations validate the framework's efficacy. The results demonstrate that SeqTG effectively eradicates late-stage bloat, achieving state-of-the-art test suite compactness and strict constraint adherence.
	\end{abstract}
	
	
	\section{Introduction}
	\label{sec:intro}
	
	Software quality assurance is increasingly critical as modern systems grow in complexity \citep{pressman2014}. Conventional verification techniques primarily target functional correctness and often fail to expose faults triggered only when multiple configuration parameters interact \citep{nie2011, cohen2004}. Since empirical evidence shows that interaction-triggered faults constitute a major portion of field failures, systematic techniques for managing combinatorial explosion are indispensable \citep{kuhn2004}.
	
	Combinatorial testing (CT) addresses this challenge by generating a test suite that covers $t$-way parameter interactions \citep{nie2011}. Pairwise testing (i.e., $t=2$) is particularly popular in industry, balancing high fault-detection rates with test suite compactness \citep{1997The}. This compactness is vital in modern automated environments like regression testing \citep{Saha2023} and Continuous Integration/Continuous Deployment (CI/CD) pipelines \citep{Otterklau2022}. Because these test suites are executed repeatedly over the software lifecycle, accumulated computational and time costs are high; thus, eliminating even a single redundant test case yields substantial savings, making the pursuit of near-optimal suite sizes a high priority.
	
\begin{figure*}[!t]
		\centering
		\includegraphics[width=0.9\textwidth]{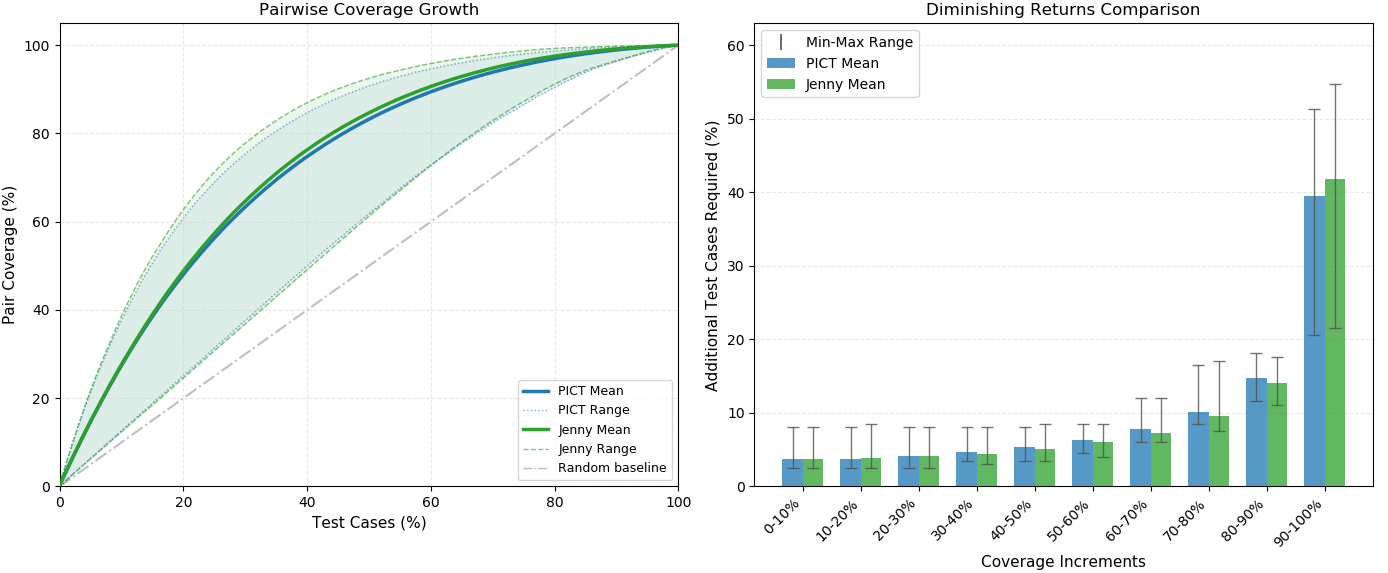}
		\caption{The efficiency curve and diminishing returns of greedy combinatorial testing algorithms. The data is derived from 100 independent runs of PICT and Jenny on randomly generated system configurations (with the number of factors and levels uniformly ranging from 5 to 30). The left panel illustrates the mean pairwise coverage growth and its variance, showing rapid initial coverage. The right panel quantifies the inefficiency in the later stages, demonstrating that covering the final 10\% of interactions (90-100\%) disproportionately consumes roughly 40\% of the total generated test cases.}
		\label{fig:motivation}
	\end{figure*}

	However, generating a minimal-size Covering Array (CA) under complex industrial constraints remains an NP-hard challenge. Existing greedy algorithms (e.g., AETG, IPOG, PICT) \citep{1997The,2007IPOG,Czerwonka2006} are highly scalable but suffer from inherent shortsightedness. By making locally optimal decisions without global foresight, they inevitably leave behind a fragmented tail of disjointed, ``isolated'' pairs. Packing these unaligned interactions late in the process is highly inefficient, ultimately bloating the suite with redundant, low-yield tests. Alternatively, while metaheuristics (e.g., GA, PSO) \citep{Shiba2004,Ahmed2012} attempt to alleviate this via global search, they lack determinism and formal optimality guarantees. This performance bottleneck is empirically demonstrated in Fig.~\ref{fig:motivation}. Across 100 randomly generated configurations, industry-standard greedy algorithms (PICT and Jenny) exhibit severe ``diminishing returns.'' While they rapidly cover the initial majority of pairs, covering the final 10\% of interactions disproportionately consumes approximately 40\% of the total generated test cases. This myopic search for the remaining ``difficult'' pairs inevitably bloats the test suite, rendering a uniform greedy strategy inherently suboptimal.
	
	To rigorously address this inefficiency, we first establish a \textbf{monolithic ILP formulation} that models the entire test generation as a single global optimization task. Although it theoretically guarantees a minimal test suite and provides precise constraint handling, this monolithic approach suffers from crippling combinatorial explosion, rendering it computationally intractable for large-scale industrial systems. Instead of solving one prohibitive NP-hard model, we propose \textbf{SeqTG} which decomposes the problem into a sequence of lightweight ILP subproblems. It employs a ``warm-start'' strategy: utilizing a fast greedy baseline to swiftly clear the ``easy'' initial phase, while reserving exact sequential ILP exclusively for the inefficient, difficult tail. By generating a single mathematically optimized test case at each step to cover the remaining interactions, SeqTG bridges the scalability of heuristics with the rigorous quality of exact optimization.

	The specific contributions of this paper are summarized as follows:
	
	\begin{itemize}
		\item \textbf{A Monolithic ILP Baseline.} We establish a rigorous mathematical programming model for constrained covering array generation. This formulation guarantees global optimality and provides a unified method for handling complex inclusionary and exclusionary constraints, serving as a robust theoretical benchmark for our study.
		
		\item \textbf{A Scalable Sequential Framework.} We propose a decomposition strategy that transforms the intractable monolithic problem into a sequence of solvable ILP instances. Central to this is a \textit{Warm-Start} mechanism: we leverage fast greedy tools (e.g., PICT) to rapidly cover easy pairs, reserving our rigorous sequential ILP specifically for the difficult, fragmented tail. This hybrid approach achieves greedy-like scalability while retaining the monolithic model's precision.
		
		\item \textbf{Constraint-First Preprocessing and Global Minimization.} We develop Greedy Constraint Partitioning (GCP) to efficiently handle must-include requirements and apply a final Set Covering formulation to eliminate redundancy, ensuring the compactness of the final suite.
		
		\item \textbf{Extensive Evaluation.} We extensively evaluate SeqTG across an industrial 5G case study, standard benchmarks, and 200 large-scale configurations. Results show that SeqTG effectively addresses complex, large-scale problems and consistently generated significantly more compact test suites than existing tools, securing top-rank compactness in 94\% of unconstrained and 99\% of heavily constrained instances.	
	\end{itemize}
	
	The remainder of this paper is organized as follows. Section~\ref{sec:framework} details the mathematical formulations of both the monolithic ILP baseline and the sequential SeqTG framework. Section~\ref{sec:experiments} reports the experimental evaluation. Section~\ref{sec:discussion} discusses limitations and future directions, and Section~\ref{sec:conclusion} concludes the paper.

	\section{The SeqTG Framework}
	\label{sec:framework}
	The core challenge in generating an optimal Covering Array (CA) lies in its NP-hard nature. While theoretically elegant, formulating the entire problem as a single, large-scale optimization model—a monolithic approach—quickly becomes computationally intractable. This section first explores such monolithic formulations to establish a theoretical baseline and highlight their inherent scalability issues, thereby motivating the sequential, decomposed strategy employed by our proposed framework, SeqTG.
	
	\subsection{Monolithic Optimization Approaches}
	\label{sec:monolithic}
	A monolithic approach attempts to find a globally optimal CA of a fixed size $m$ by defining and solving for all test cases simultaneously. To establish a theoretical benchmark, we construct a monolithic Integer Linear Programming (ILP) formulation that reformulates pairwise CT as a global optimization problem by directly maximizing pair coverage. This rigorous approach guarantees globally minimal test suites with formal optimality certificates and natively encodes complex constraints. However, as detailed below, its formulation suffers from severe scalability limitations.
	
	\subsubsection{ILP Formulation for Maximizing Coverage}
	To construct a dense and efficient CA, the direct mathematical objective is to maximize the number of unique pairs covered by the test suite. Let $x_{cij}$ be a binary variable that is 1 if test case $c \in \{1, \dots, m\}$ selects level $j$ for factor $i$. Let $U$ be the set of all possible pairs. We introduce a binary variable $p_u$ for each pair $u \in U$, which is 1 if the pair is covered by at least one test case.
	
	\begin{align} 
	\max_{\boldsymbol{x},\boldsymbol{p},\boldsymbol{q}} \quad & \sum_{u \in U} p_u \label{eq:mono_obj_coverage} \\
	\text{s.t.} \quad & \sum_{j=1}^{l_i} x_{cij} = 1, \quad \forall c, i \label{eq:mono_assign_2} \\
	& q_{cu} \ge x_{ci,a} + x_{cj,b} - 1, \quad \forall c, \forall u=((i,a),(j,b)) \label{eq:mono_q_def} \\
	& p_u \le \sum_{c=1}^{m} q_{cu}, \quad \forall u \in U \label{eq:mono_p_def}
	\end{align}
	
	Here, an intermediate variable $q_{cu}$ indicates if test case $c$ covers pair $u$. Constraint~\eqref{eq:mono_assign_2} ensures each test case selects exactly one level per factor. Constraint~\eqref{eq:mono_q_def} links the factor selections to the coverage, while Constraint~\eqref{eq:mono_p_def} models the logical OR, ensuring $p_u$ is 1 if any $q_{cu}$ is 1.
	
	To handle real-world restrictions, this formulation seamlessly incorporates linear constraints \citep{2006Production, 2009IBM}. Exclusionary constraints (forbidden combinations of $p$ levels) are modeled directly as linear inequalities for every test case $c$: $\sum_{s=1}^{p} x_{c i^{(s)} j^{(s)}} \leq p-1$. For inclusionary constraints (must-include combinations of $q$ levels), we introduce an auxiliary binary variable $u_c$ indicating if test case $c$ satisfies the requirement, enforced via:
	\begin{align}
	u_c &\leq x_{c i^{(s)} j^{(s)}}, \quad \forall s \in \{1, \dots, q\}, \\
	u_c &\geq \sum_{s=1}^{q} x_{c i^{(s)} j^{(s)}} - (q-1).
	\end{align}
	A global constraint $\sum_{c=1}^{m} u_c \geq 1$ then guarantees the mandatory combination appears at least once in the suite.
	
	\subsubsection{The Scalability Barrier}
	While this monolithic formulation is theoretically sound and provides global optimality with robust constraint support, its practical application is crippled by a catastrophic scalability problem. The number of variables and constraints grows polynomially with the test suite size ($m$) and the number of factors ($n$), leading to a combinatorial explosion. For most industrial-scale problems (e.g., $n > 20$), solving such a model is computationally infeasible, as confirmed by timeouts in our experiments. Complex constraints exacerbate fragmentation of the feasible space, further hindering solvers. This necessitates a more practical approach, motivating SeqTG's sequential decomposition.
	
	\subsection{From Monolithic to Sequential Optimization}
	\label{sec:sequential_model}
	To overcome the scalability barrier of the monolithic models, SeqTG fundamentally reframes the problem. Instead of solving one large, NP-hard problem, it decomposes it into a sequence of smaller, tractable ILP problems. This sequential approach marries the rigorous optimization power of ILP with the scalability of a greedy strategy. At each step, SeqTG generates a \textbf{single, highly optimized test case} designed to contribute maximally to the overall coverage goal, while inheriting the robust constraint linearization techniques from the monolithic method but applying them per-step for efficiency.
	
	This decomposition leads to a dramatic simplification of the optimization model. Instead of solving for the entire suite simultaneously, the model restricts its focus to the single test case currently being generated. Consequently, we directly reuse the fundamental factor assignment variables $x_{i,j}$ (as defined previously) for this current test case. Let $U_{c-1}$ be the set of pairs that remain uncovered before generating the $c$-th test case. To track the new coverage, we only need to introduce the following auxiliary variable:
	
	\noindent\textbf{Coverage Variable:}
	\begin{itemize}
		\item $p_{u} \in \{0, 1\}$: 1 if an uncovered pair $u \in U_{c-1}$ is covered by the \textit{new test case}; 0 otherwise.
	\end{itemize}
	
	\noindent\textbf{Objective Function Construction:}
	Our objective function is designed to prioritize more valuable interactions. The most basic goal is to cover as many new pairs as possible. However, not all pairs are created equal. Some pairs, particularly those involving factors with many levels, are statistically harder to cover by chance. To guide the solver towards these "difficult" pairs first, we introduce a weight $w_u$ for each pair $u$. This leads to our primary objective: maximizing the \textbf{weighted coverage}.
	\begin{equation}
	\label{eq:weighted_coverage_obj}
	\max_{\boldsymbol{x},\boldsymbol{p}} \quad \sum_{u \in U_{c-1}} w_u \cdot p_u
	\end{equation}
	A common and effective weighting strategy, which we adopt, is to set the weight of a pair $u=((i,a),(j,b))$ as the product of the number of levels of its constituent factors, i.e., $w_u = l_i \cdot l_j$. This heuristic intelligently prioritizes covering pairs from factors with larger domains, often leading to a more compact final test suite. This aligns with the concept of the "big rocks first" principle. Imagine the challenge of packing rocks of all sizes efficiently into a container. A naive strategy of adding smaller pebbles first could leave no space for larger rocks, leading to ineffective space utilization. The most effective approach is to place the biggest rocks first, and then fill the remaining voids with smaller ones. In combinatorial testing, pairs from factors with many levels are the "big rocks"—they are statistically harder to cover. The weight $w_u$ operationalizes this principle, effectively prioritizing these "big rock" pairs and leading to a more compact final test suite.
	
	\noindent\textbf{Constraints:}
	\begin{enumerate}
		\item \textbf{Factor Assignment:} Every factor must be assigned exactly one level.
		\begin{equation}
		\label{eq:seq_assign}
		\sum_{j=0}^{l_i-1} x_{i,j} = 1, \quad \forall i \in \{0, \dots, n-1\}
		\end{equation}
		
		\item \textbf{Pair Coverage Association:} These constraints link the test case structure ($x$ variables) to the pair coverage indicators ($p$ variables). For each uncovered pair $u = ((i, a), (j, b))$, the logical AND condition ($p_u=1 \iff x_{i,a}=1 \land x_{j,b}=1$) is robustly linearized with:
		\begin{align}
		\label{eq:seq_link_p_1_simplified}
		p_{u} &\leq x_{i,a} \\
		\label{eq:seq_link_p_2_simplified}
		p_{u} &\leq x_{j,b}
		\end{align}
	\end{enumerate}
	Exclusionary and inclusionary constraints from the monolithic baseline are adapted here per-step, using the same linearization but on a significantly reduced scale. This sequential model is drastically smaller and faster to solve than its monolithic counterparts, yet it leverages the full power of ILP at each step to make a locally optimal decision.

	\subsection{Constraint Handling in the Sequential Model}
	A key advantage of SeqTG is the native and rigorous handling of complex real-world constraints within its sequential ILP model, building on the exact linearization techniques established in our monolithic approach.
	
	Exclusionary constraints (forbidden combinations of $p$ levels) are enforced via $\sum_{s=1}^p x_{i^{(s)},j^{(s)}} \leq p-1$. For inclusionary constraints, the Greedy Constraint Partitioning (GCP) preprocessor groups compatible must-includes to minimize dedicated test cases, using graph-based incompatibility detection (conflicts or exclusion violations).
	
	Before describing the grouping algorithm, it is crucial to formally define what makes a set of must-include constraints "compatible". Two or more must-include constraints are considered \textbf{compatible} if they can be simultaneously satisfied within a single, valid test case. Conversely, they are \textbf{incompatible} if either of the following conditions is met:
	\begin{itemize}
		\item \textbf{Conflicting Factor Assignment:} The constraints attempt to assign different levels to the same factor. For example, a constraint requiring `(OS=Windows)` and another requiring `(OS=Linux)` are incompatible because the factor `OS` cannot hold both values at once.
		\item \textbf{Violation of Exclusionary Constraints:} The union of all factor-level assignments required by the constraints forms a combination that is explicitly forbidden by an exclusionary constraint in $C_{avoid}$. For instance, if the system has a rule `IF A=1 THEN B!=2`, the must-include constraints `\{A=1\}` and `\{B=2\}` are incompatible.
	\end{itemize}
	The incompatibility graph $G$ used in Algorithm~\ref{alg:gcp} is constructed based on this definition, where an edge connects any two constraints that are found to be incompatible. The core idea is to group multiple compatible must-include constraints so they can be satisfied by a single test case. We model this as a graph problem where vertices are constraints and an edge connects incompatible constraints. Algorithm~\ref{alg:gcp} provides an effective heuristic, GCP, to find compatible partitions $P = \{S_1, \dots, S_k\}$. For each compatible group $S_k$ from GCP, SeqTG solves the sequential ILP model with an additional set of equality constraints. For a group $S_k$ requiring levels $\{(i^{(1)}, j^{(1)}), \dots, (i^{(q)}, j^{(q)}) \}\}$, the following constraints are added:
	\begin{equation} 
	\label{eq:inclusion}
	x_{i^{(s)}, j^{(s)}} = 1, \quad \forall s \in \{1, \dots, q\}
	\end{equation}
	This ensures the generated test case satisfies all constraints in $S_k$ while still being optimized to cover other new interactions.
	
		\begin{algorithm}[h]
		\caption{Greedy Constraint Partitioning (GCP)}
		\label{alg:gcp}
		\begin{algorithmic}[1]
			\State \textbf{Input:} Set of must-include constraints $C_{must}$
			\State \textbf{Output:} A partition of $C_{must}$ into compatible subsets $P$
			\State Build the incompatibility graph $G = (V, E)$ from $C_{must}$
			\State Sort vertices in $V$ in descending order of degree into a list $V_{sorted}$
			\State Initialize $P \leftarrow \emptyset$, $k \leftarrow 0$
			\State \textbf{while} $V_{sorted}$ is not empty \textbf{do}
			\State \quad $k \leftarrow k + 1$; $S_k \leftarrow \emptyset$
			\State \quad Take the first constraint $c_{first}$ from $V_{sorted}$
			\State \quad Add $c_{first}$ to $S_k$ and remove it from $V_{sorted}$
			\State \quad \textbf{for each} constraint $c_j$ in a copy of $V_{sorted}$ \textbf{do}
			\State \quad \quad \textbf{if} $c_j$ is compatible with all constraints in $S_k$ \textbf{then}
			\State \quad \quad \quad Add $c_j$ to $S_k$; Remove $c_j$ from $V_{sorted}$
			\State \quad \quad \textbf{end if}
			\State \quad \textbf{end for}
			\State \quad Add $S_k$ to $P$
			\State \textbf{end while}
			\State \textbf{return} $P$
		\end{algorithmic}
	\end{algorithm}

	\subsection{Global Test Suite Minimization via Set Covering}
	The sequential generation process may produce a test suite $TS_{cand}$ that contains redundancy. To address this, SeqTG incorporates a final optimization phase using the classic \textbf{Set Covering} model. This step finds the smallest possible subset of $TS_{cand}$ that still covers all required interactions.
	
	\noindent\textbf{Decision Variables:}
	\begin{itemize}
		\item $z_{c} \in \{0, 1\}$: 1 if test case $tc_c \in TS_{cand}$ is selected for the final suite; 0 otherwise.
	\end{itemize}
	
	\noindent\textbf{Set Covering ILP Formulation:}
	The objective is to minimize the total number of selected test cases:
	\begin{equation}
	\label{eq:sc_obj}
	\min_{\boldsymbol{z}} \quad \sum_{c=1}^{m} z_c
	\end{equation}
	subject to the coverage constraint: for every required interaction $u$, at least one selected test case must cover it.
	\begin{equation}
	\label{eq:sc_constraint}
	\sum_{tc_c \in C_u} z_c \geq 1, \quad \forall u \in P_{all}
	\end{equation}
	where $C_u \subseteq TS_{cand}$ is the subset of candidate test cases that cover interaction $u$. This model elegantly identifies and discards every non-essential test case, achieving near-global optimality without the full monolithic scale.
	
	\subsection{The Complete SeqTG Algorithm}
	\label{sec:complete_algorithm}
	
	The complete SeqTG framework, detailed in Algorithm~\ref{alg:seqtg}, integrates the previously defined components into a structured, four-stage process. This hybrid architecture is designed to capitalize on the strengths of both greedy heuristics and formal optimization, directly addressing the diminishing returns problem shown in Fig.~\ref{fig:motivation}.
	
	The process is as follows:
	\begin{itemize}
		\item \textbf{Warm-Start:} The algorithm is initialized with a test suite from a fast greedy tool (e.g., PICT). The `ApplyWarmStart` function filters this suite against exclusionary constraints, calculates the initial pair coverage, and identifies which `must-include` constraints are already satisfied. This step rapidly reduces the problem space.
		\item \textbf{Targeted Constraint Generation:} The `HandleMustInclude` function addresses any `must-include` constraints that remain unsatisfied. It uses Greedy Constraint Partitioning (GCP) and the sequential ILP model to generate the minimum number of test cases required to fulfill these constraints.
		\item \textbf{Incremental Coverage Generation:} The main `while` loop then iteratively calls the `GenerateSingleCase` function to cover all remaining interactions until 100\% coverage is achieved.
		\item \textbf{Global Suite Minimization:} Finally, the `MinimizeSuite` function uses a Set Covering model to analyze the entire candidate suite and eliminate any redundant test cases, yielding a compact and complete final suite.
	\end{itemize}
	This staged decomposition allows SeqTG to maintain the scalability of greedy methods while leveraging the optimization power of ILP for superior test suite quality.
	
		\begin{algorithm}[h]
		\caption{The SeqTG Framework}
		\label{alg:seqtg}
		\begin{algorithmic}[1]
			\State \textbf{Input:}
			\State \quad $S_{factors}, t, C_{must}, C_{avoid}$ 
			\State \quad $TS_{init}$ 
			\State \textbf{Output:} A minimized test suite $TS_{final}$.
			
			\State \vspace{0.5em}
			\State $U_{all} \leftarrow \text{GetAllAchievableInteractions}(S_{factors}, t, C_{avoid})$.
			\State Initialize $TS_{cand} \leftarrow \emptyset$, $U_{covered} \leftarrow \emptyset$.
			
			\State \vspace{0.5em}
			\State \textit{// Phase 0: Warm-Start with a greedy solution}
			\State $(TS_{cand}, U_{covered}, C_{must}) \leftarrow \text{ApplyWarmStart}(TS_{init}, C_{avoid}, C_{must}, TS_{cand}, U_{covered})$.
			
			\State \vspace{0.5em}
			\State \textit{// Phase 1: Handle remaining must-include constraints}
			\State $(TS_{cand}, U_{covered}) \leftarrow \text{HandleMustInclude}(C_{must}, C_{avoid}, U_{all} \setminus U_{covered}, TS_{cand}, U_{covered})$.
			
		    \State \vspace{0.5em}
			\State \textit{// Phase 2: Incrementally cover remaining interactions}
			\State \textbf{while} $U_{covered} \neq U_{all}$ \textbf{do}
			\State \quad $tc_{new} \leftarrow \text{GenerateSingleCase}(U_{all} \setminus U_{covered}, \emptyset, C_{avoid})$.
			\State \quad \textbf{if} $tc_{new}$ is null \textbf{then break} \textbf{end if}
			\State \quad Add $tc_{new}$ to $TS_{cand}$; Update $U_{covered}$.
			\State \textbf{end while}
			
			\State \vspace{0.5em}
			\State \textit{// Phase 3: Globally minimize the suite}
			\State $TS_{final} \leftarrow \text{MinimizeSuite}(TS_{cand}, U_{all})$.
			
			\State \vspace{0.5em}
			\State \textbf{return} $TS_{final}$
		\end{algorithmic}
	\end{algorithm}

	\section{Experiments and Results}
	\label{sec:experiments}
	
	To comprehensively evaluate the effectiveness and scalability of our proposed SeqTG framework, we conducted a series of experiments benchmarked against leading combinatorial testing tools. Our evaluation compares SeqTG primarily with \textbf{PICT}, a highly efficient greedy algorithm, as well as a wide array of state-of-the-art academic and industrial tools where published results are available, including \textbf{AETG}, \textbf{IPOG}, and \textbf{Jenny}. We also include the aforementioned \textbf{Monolithic ILP} formulation as a baseline to strictly highlight the scalability challenges inherent in direct exact methods.
	
	All new experiments were executed on a workstation with an Intel Core i9-13900H CPU and 32 GB of RAM, using CPLEX as the underlying ILP solver. The primary evaluation metrics are the final \textbf{Test Suite Size}, which reflects testing economy, and the \textbf{Generation Time (s)}, which measures computational performance. The evaluation is structured into three parts: a broad comparison on standard benchmarks, a scalability analysis on unconstrained problems, and a performance analysis on constrained problems.
	
	\subsection{Industrial Case Study: 5G Baseband Configuration Testing}
	To further validate the practical utility of SeqTG in a realistic industrial context, we applied the proposed method to the configuration testing of a 5G Baseband Unit (BBU). This specific domain is characterized by a multi-dimensional parameter space and strict operational constraints derived from both hardware limitations and 3GPP protocol standards. The System Under Test (SUT) is defined by four key configuration parameters: Modulation, Bandwidth, MIMO Mode, and Coding Rate. As outlined in Table~\ref{tab:5g_case_study}, each factor comprises four distinct levels, creating a theoretical search space of 256 distinct combinations.
	
	\begin{table}[h!]
		\centering
		\caption{Factor-Level Configuration for the 5G Baseband Unit}
		\label{tab:5g_case_study}
		\begin{tabular}{@{}lllll@{}}
			\toprule
			\textbf{Factor} & \textbf{Level 1} & \textbf{Level 2} & \textbf{Level 3} & \textbf{Level 4} \\ \midrule
			Modulation & QPSK & 16-QAM & 64-QAM & 256-QAM \\
			Bandwidth & 20 MHz & 50 MHz & 100 MHz & 200 MHz \\
			MIMO Mode & SU-MIMO & MU-MIMO & Massive MIMO & No MIMO \\
			Coding Rate & 1/3 & 1/2 & 3/4 & 5/6 \\ \bottomrule
		\end{tabular}
	\end{table}

	To simulate the rigorous requirements of carrier acceptance testing, we incorporated two representative constraints into the generation process. The first is a mandatory inclusionary constraint designed to verify a specific high-throughput profile; the test suite must explicitly include at least one test case containing the tuple \{\texttt{Modulation=256-QAM}, \texttt{Bandwidth=200 MHz}, \texttt{MIMO Mode=MU-MIMO}\}. The second is an exclusionary constraint addressing hardware signal processing limitations at the network edge. Specifically, the simplest modulation scheme is physically incompatible with the widest bandwidth, rendering the combination \{\texttt{Modulation=QPSK}, \texttt{Bandwidth=200 MHz}\} strictly forbidden in any valid test case.
	
	The test suite generated by SeqTG, comprising 21 optimized test cases, is presented in Table~\ref{tab:5g_generated_suite}. A detailed inspection of the results demonstrates the tool's capability to navigate the constrained space effectively while ensuring full pairwise coverage of the remaining valid interactions. The solver prioritized the inclusionary constraint immediately; the mandatory combination of 256-QAM, 200 MHz, and MU-MIMO is successfully generated in Test Case 1 (\texttt{TC1}). Notably, the algorithm selected \texttt{Coding Rate=1/3} to complete this tuple, ensuring the test case contributes effectively to the overall coverage rather than serving solely as a constraint patch.
	
	\begin{table*}[!t]
		\centering
		\caption{Generated Test Suite for the 5G Baseband Case Study (Size=21)}
		\label{tab:5g_generated_suite}
			\begin{tabular}{@{}cllll@{}}
				\toprule
				\textbf{TC} & \textbf{Modulation} & \textbf{Bandwidth} & \textbf{MIMO Mode} & \textbf{Coding Rate} \\ \midrule
				1           & 256-QAM             & 200 MHz            & MU-MIMO            & 1/3                  \\
				2           & QPSK                & 20 MHz             & SU-MIMO            & 1/3                  \\
				3           & 256-QAM             & 100 MHz            & No MIMO            & 1/2                  \\
				4           & 16-QAM              & 200 MHz            & Massive MIMO       & 5/6                  \\
				5           & 64-QAM              & 20 MHz             & No MIMO            & 3/4                  \\
				6           & QPSK                & 100 MHz            & MU-MIMO            & 5/6                  \\
				7           & 16-QAM              & 50 MHz             & SU-MIMO            & 3/4                  \\
				8           & 64-QAM              & 50 MHz             & Massive MIMO       & 1/3                  \\
				9           & 64-QAM              & 200 MHz            & SU-MIMO            & 1/2                  \\
				10          & 16-QAM              & 20 MHz             & MU-MIMO            & 1/2                  \\
				11          & QPSK                & 50 MHz             & Massive MIMO       & 1/2                  \\
				12          & 256-QAM             & 20 MHz             & Massive MIMO       & 3/4                  \\
				13          & 256-QAM             & 50 MHz             & No MIMO            & 5/6                  \\
				14          & 16-QAM              & 100 MHz            & No MIMO            & 1/3                  \\
				15          & 64-QAM              & 100 MHz            & MU-MIMO            & 3/4                  \\
				16          & 256-QAM             & 100 MHz            & SU-MIMO            & 5/6                  \\
				17          & 64-QAM              & 200 MHz            & No MIMO            & 5/6                  \\
				18          & QPSK                & 20 MHz             & No MIMO            & 5/6                  \\
				19          & QPSK                & 100 MHz            & Massive MIMO       & 3/4                  \\
				20          & 16-QAM              & 50 MHz             & MU-MIMO            & 3/4                  \\
				21          & 16-QAM              & 200 MHz            & SU-MIMO            & 3/4                  \\ \bottomrule
			\end{tabular}%
	\end{table*}
	
	Furthermore, the exclusionary constraint is rigorously maintained throughout the suite, proving the robustness of the constraint handling mechanism. The forbidden interaction is entirely absent from the generated set. For instance, wherever \texttt{QPSK} appears (specifically in TC 2, 6, 11, 18, and 19), it is exclusively paired with safe bandwidth settings (20, 50, or 100 MHz). Conversely, all instances involving the high-bandwidth setting of \texttt{200 MHz} (TC 1, 4, 9, 17, and 21) are correctly matched with higher-order modulation schemes such as 16-QAM, 64-QAM, or 256-QAM. This consistent adherence to complex logical rules confirms that SeqTG generates a valid, operational test suite capable of direct deployment in 5G verification environments without manual intervention.

	\subsection{Comparison on Standard Benchmarks and Weight Ablation}
	\label{sec:benchmark_comparison}
	
	To situate SeqTG within the broader academic landscape, we conducted a comprehensive comparison against a spectrum of state-of-the-art tools using widely accepted standard benchmarks. Table~\ref{tab:comparison_results} summarizes the results on specific arrays, while Table~\ref{tab:param_comparison} illustrates scalability across a 10-parameter system with variable levels ($V$). We compared SeqTG with greedy and deterministic algorithms (AETG \citep{1997The}, mAETG \citep{mAETG_ref}, IPOG \citep{2007IPOG}, Jenny \citep{Jenny_ref}, TConfig \citep{TConfig_ref}, PICT \citep{Czerwonka2006}) and metaheuristic approaches (GA \citep{Shiba2004}, ACA \citep{ACA_ref}, PSO \citep{Ahmed2012}). Importantly, to isolate the rigorous optimization capability of our mathematical model, we disabled the Warm-Start mechanism in SeqTG for these smaller-scale experiments.
	
	Furthermore, to directly validate the effectiveness of our weighted objective function (Eq.~\ref{eq:weighted_coverage_obj}), we introduce an unweighted ablation variant, \textbf{SeqTG (nw)}, which assigns an equal weight ($w_u = 1$) to all pairs.
	
	On fixed-level (homogeneous) covering arrays, the performance differences are negligible for small instances (e.g., $3^3, 3^4$). However, as the problem scale increases (Table~\ref{tab:param_comparison}), the limitations of traditional greedy strategies become evident. For the challenging $CA(10^{10})$ benchmark, greedy tools like IPOG and PICT generated test suites of sizes 176 and 170, respectively. In contrast, SeqTG produced a suite of only \textbf{151} test cases, reducing the size by \textbf{14.2\%} compared to IPOG and \textbf{11.2\%} compared to PICT.
	
	The advantages of SeqTG are especially pronounced in mixed-level (heterogeneous) systems, which present a unique challenge as parameters with larger domains are statistically harder to cover. For the highly heterogeneous benchmark $MCA(10^1 9^1 8^1 7^1 6^1 5^1 4^1 3^1 2^1)$, SeqTG achieved a size of \textbf{91}, outperforming PICT (101) and PSO (97). Crucially, the unweighted variant \textbf{SeqTG (nw)} required \textbf{95} test cases for this same configuration. This performance gap highlights the importance of the ``big rocks first'' principle. Without size-based weights, the ILP solver naturally picks the ``low-hanging fruit''---numerous easy pairs from factors with fewer levels---leaving behind a fragmented tail of ``difficult'' pairs that bloats the final suite. Prioritizing pairs by their domain size effectively eradicates this inefficiency.
	
	Finally, while the monolithic ILP baseline provides global optimality for extremely small instances, it failed to produce a solution within the 3600-second timeout for the majority of these benchmarks (e.g., $5^{10}$). SeqTG bridges this gap by applying rigorous constraints sequentially, achieving significantly better results on large-scale benchmarks without hitting the scalability wall of monolithic approaches.

\begin{table*}[!t]
	\centering
	\caption{Comparison with Existing Tools on Standard Benchmarks (t=2)}
	\label{tab:comparison_results}
	\resizebox{\linewidth}{!}{
		\begin{tabular}{lccccccccccc}
			\toprule
			\textbf{Configurations} & \textbf{AETG} & \textbf{mAETG} & \textbf{GA} & \textbf{ACA} & \textbf{IPOG} & \textbf{Jenny} & \textbf{TConfig} & \textbf{PICT} & \textbf{PSO} & \textbf{SeqTG (nw)}& \textbf{SeqTG} \\ 
			\midrule
			$CA(3^3)$ & NA & NA & NA & NA & 11 & 9 & 10 & 10 & 9  & 10 & 10 \\ 
			$CA(3^4)$ & 9 & 9 & 9 & 9 & 12 & 13 & 10 & 13 & 9& 9 &  9 \\ 
			$CA(3^{10})$ & 15 & 17 & 17 & 17 & 12 & 20 & 20 & 20 & 17 & 15 &  15 \\ 
			$CA(5^{10})$ & NA & NA & NA & NA & 50 & 45 & 48 & 47 & 45& 42  &  44 \\ 
			$CA(10^{10})$ & NA & NA & 157 & 159 & 176 & 157 & 170 & 170 & 170 & 152 &  151 \\ 
			$MCA(5^1 3^8 2^2)$ & 19 & 20 & 15 & 16 & 19 & 41 & 22 & 21 & 21& 20 & 20  \\ 
			$MCA(6^1 5^1 4^6 3^8 2^3)$ & 34 & 35 & 33 & 32 & 36 & 31 & 33 & 38 & 39 & 39 &  35 \\ 
			$MCA(7^1 6^1 5^1 4^6 3^8 2^3)$ & 45 & 44 & 42 & 42 & 44 & 51 & 49 & 46 & 49 & 53 & 45 \\ 
			$MCA(10^1 9^1 8^1 7^1 6^1 5^1 4^1 3^1 2^1)$ & NA & NA & NA & NA & 91 & 98 & 92 & 101 & 97& 95 &  91 \\ 
			\bottomrule
		\end{tabular}
	} 
\end{table*}

	\begin{table}[htbp]
		\centering
		\caption{Comparison on 10-Parameter System with Variable Levels (V)}
		\label{tab:param_comparison}
		\begin{tabular}{ccccccc}
			\toprule
			V & IPOG & Jenny & TConfig & PICT & PSO & SeqTG\\ 
			\midrule
			3 & 20 & 19 & 17 & 18 & 17  &  15 \\ 
			4 & 31 & 30 & 31 & 31 & 29  & 30 \\ 
			5 & 50 & 45 & 48 & 47 & 45  & 44 \\ 
			6 & 68 & 62 & 64 & 66 & 62 & 59 \\ 
			7 & 90 & 83 & 85 & 88 & 81 & 79 \\ 
			8 & 117 & 104 & 114 & 112 & 109  &98 \\ 
			9 & 142 & 129 & 139 & 139 & 139 & 122 \\ 
			10 & 176 & 157 & 170 & 170 & 170 &151 \\ 
			\bottomrule
		\end{tabular}
	\end{table}
	
	\subsection{Performance on Large-Scale Unconstrained Problems}
	\label{sec:unconstrained_perf}

\begin{figure*}[!t]
		\centering
		\includegraphics[width=\textwidth]{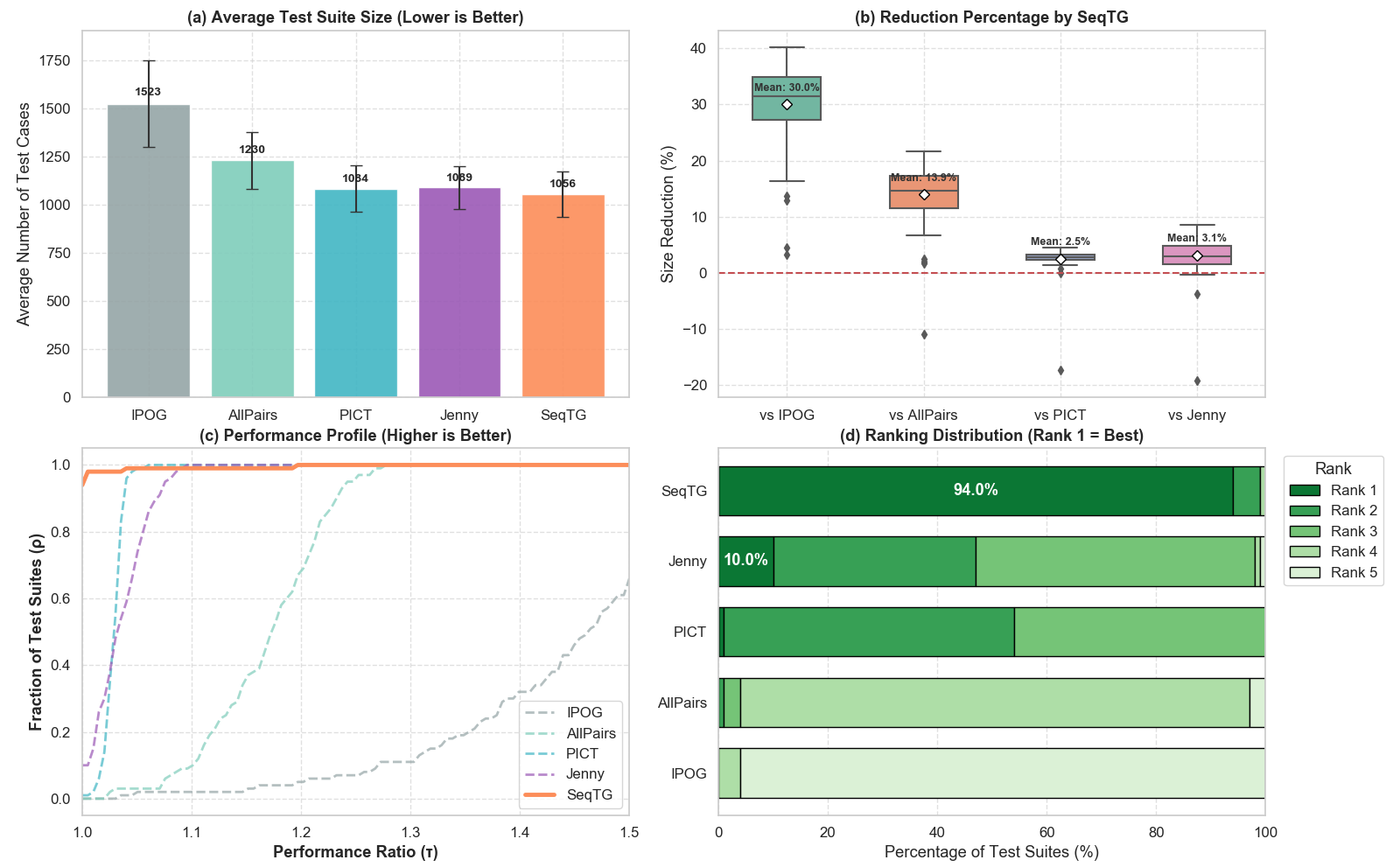}
		\caption{Comprehensive performance comparison of SeqTG against IPO, AllPairs, PICT, and Jenny across 100 unconstrained test configurations. (a) Average test suite size with standard deviation. (b) Box plots showing the percentage reduction in test suite size achieved by SeqTG relative to other methods. (c) Performance profiles illustrating the fraction of problems solved within a given performance ratio. (d) Stacked bar chart showing the ranking distribution for each method (Rank 1 is best).}
		\label{fig:performance_analysis}
	\end{figure*}

	In order to test SeqTG's scalability and predictability across large-scale examples, we ran experiments under 100 arbitrary (randomly created) system configurations. These configurations are aimed at capturing the large search spaces of complex industrial environments, which are composed of 30 independent parameters in variable levels (randomly distributed between 2 and 30). This extensive dataset serves to test how reliably the algorithms maintain high-quality outputs across unknown, high-dimensional spaces. We benchmarked SeqTG against four widely used tools: \textbf{IPO}, \textbf{AllPairs}, \textbf{PICT}, and \textbf{Jenny}. The detailed statistical results are visualized in Fig.~\ref{fig:performance_analysis}.
	
	We first analyze the absolute testing economy on these large-scale datasets, as presented in Fig.~\ref{fig:performance_analysis}(a). SeqTG consistently produced the most compact test suites, achieving the lowest average size of \textbf{1056}. That is a huge step ahead of deterministic greedy algorithms, and it has a big advantage over IPOG as well as AllPairs. Relative to these perfectly optimized heuristic tools, however, SeqTG continued to dominate both PICT and Jenny. The exceptionally tight error bars associated with SeqTG indicate high stability, proving that its sequential ILP approach delivers highly predictable performance regardless of the underlying data scale.
	
	To quantify the efficiency gains on these massive configurations, Fig.~\ref{fig:performance_analysis}(b) illustrates the distribution of percentage reductions in test suite size. The box plots reveal that SeqTG achieves a massive median reduction of approximately 30\% against IPO and 14\% against AllPairs, with the entire interquartile range strictly above zero. More importantly, against the competitive baselines of PICT and Jenny, the distribution remains consistently positive, with median reductions of 2.5\% and 3.1\% respectively. The presence of outliers extending up to a 15\% reduction against PICT highlights SeqTG's robustness: in complex, large-scale instances where greedy heuristics easily get trapped in local optima, SeqTG's formal lookahead mechanism reliably uncovers deeper optimization opportunities.
	
	The predictable robustness of our approach on unseen large-scale data is further underscored by the performance profile in Fig.~\ref{fig:performance_analysis}(c), which plots the probability of a solver being within a factor $\tau$ of the best-known solution. The curve for SeqTG (solid orange) exhibits strict dominance, starting at a probability of roughly 0.94 at $\tau=1.0$. This mathematically demonstrates that as the data scale grows, SeqTG guarantees the absolute best solution in the vast majority of cases. In contrast, the curves for PICT and Jenny start significantly lower, implying that their heuristic nature leads to greater variance and suboptimal predictions on massive datasets.
	
	Finally, the ranking distribution in Fig.~\ref{fig:performance_analysis}(d) provides a definitive summary of the algorithm's capability on large-scale problems. SeqTG secured the first rank (best test suite size) in \textbf{94\%} of the 100 massive benchmark instances. The next best competitor, Jenny, achieved the top rank in only 10\% of cases (accounting for ties), while PICT and AllPairs primarily occupied the second and third ranks. This overwhelming dominance confirms that decomposing the generation process into sequential ILPs effectively resolves the scalability bottlenecks of monolithic exact models, establishing SeqTG as a highly reliable and predictable tool for large-scale combinatorial test generation.

	\subsection{Performance on Large-Scale Constrained Problems}
	\label{sec:constrained_perf}
	
	While scalability on unconstrained configurations establishes a baseline, real-world industrial software testing is invariably governed by complex logical constraints. The true test of a generator's practical utility lies in its ability to navigate highly restricted, fragmented search spaces without sacrificing testing economy. To evaluate this, we designed a rigorous stress test focusing on large-scale constrained covering array generation.
	
	It is important to emphasize that for this specific evaluation, we benchmarked SeqTG directly and exclusively against \textbf{PICT}. This decision is rooted in the empirical limitation that the vast majority of conventional combinatorial testing tools—including classic implementations of IPO, AllPairs, and standard metaheuristics (e.g., GA, PSO)—lack native, algorithmic support for complex logical constraints. Tools without an embedded constraint-solving engine typically resort to a naive ``generate-and-filter'' strategy: they generate test cases as if the space were unconstrained and then subsequently discard or attempt to patch invalid rows. In massive, highly restricted search spaces, this decoupled approach inevitably leads to catastrophic performance failures, typically resulting in solver timeouts or drastically bloated suites as the algorithm blindly struggles to find valid combinations for the final few interactions. PICT, developed by Microsoft, stands out as a notable exception. It incorporates a mature, highly optimized heuristic constraint-handling engine and is widely recognized as the industry standard for complex constrained combinatorial testing. Consequently, PICT is the only viable, competitive baseline capable of surviving this rigorous stress test.
	
\begin{figure*}[!t]
		\centering
		\includegraphics[width=\textwidth]{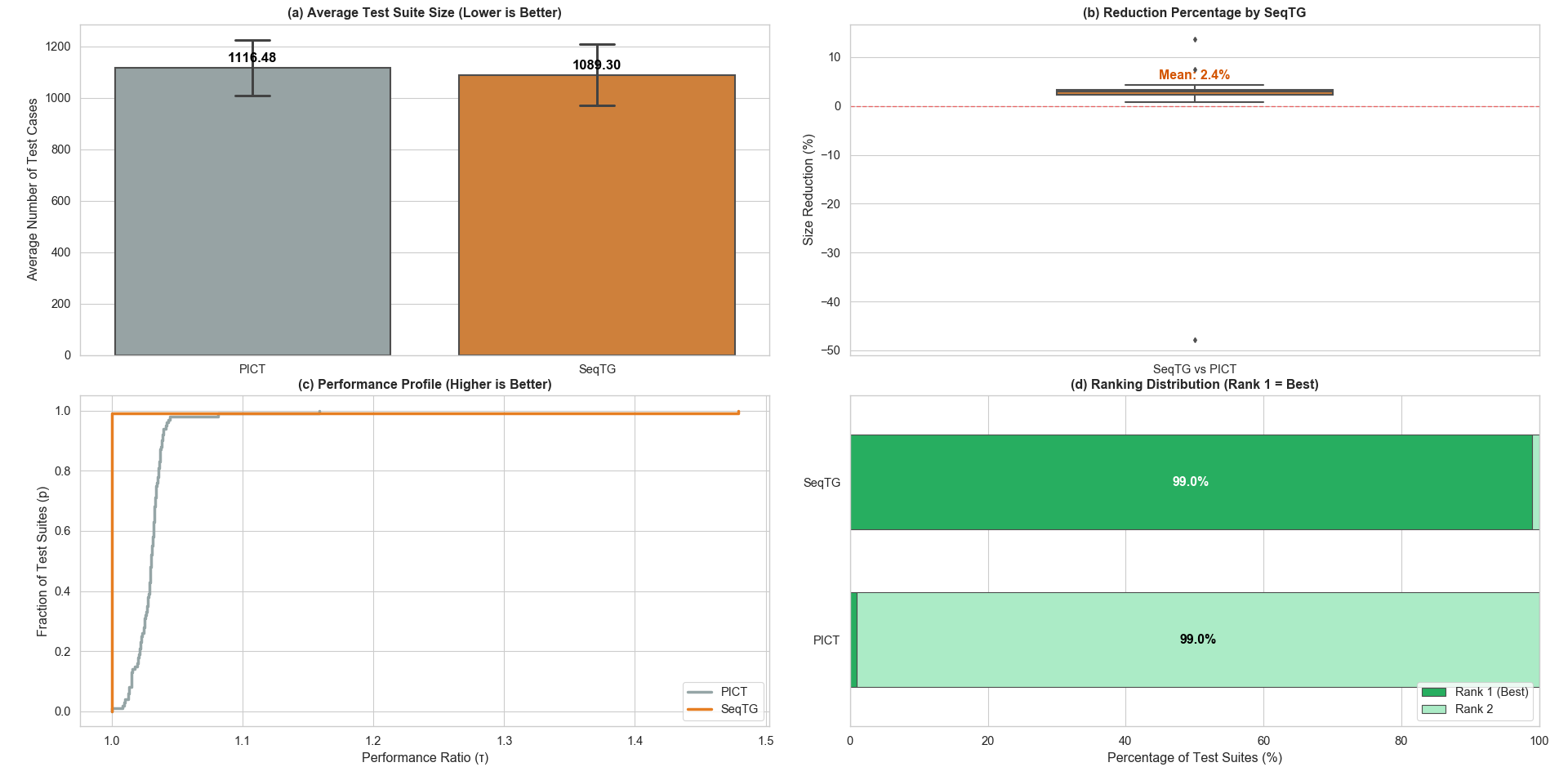}
		\caption{Comprehensive performance comparison of SeqTG against IPO, AllPairs, PICT, and Jenny across 100 constrained test configurations. (a) Average test suite size with standard deviation. (b) Box plots showing the percentage reduction in test suite size achieved by SeqTG relative to other methods. (c) Performance profiles illustrating the fraction of problems solved within a given performance ratio. (d) Stacked bar chart showing the ranking distribution for each method (Rank 1 is best).}
		\label{fig:performance_analysis_2}
	\end{figure*}

	We generated a challenging benchmark set of 100 synthetic system configurations, each featuring \textbf{30 parameters} with domain sizes randomly varying between \textbf{2 and 30}. This variation creates a massive, heterogeneous state space populated with strict validity rules. The comprehensive results comparing SeqTG and PICT are presented in Fig. \ref{fig:performance_analysis_2}.
	
	As illustrated in Fig.~\ref{fig:performance_analysis_2}(a), SeqTG maintains a clear advantage in testing economy even under heavy constraints. Across the 100 massive instances, PICT produced an average test suite size of \textbf{1116.48}, whereas SeqTG successfully compressed the average size to \textbf{1089.30}. The consistency of this efficiency is highlighted in the reduction distribution shown in Fig.~\ref{fig:performance_analysis_2}(b). The box plot reveals that the reduction percentage is strictly positive for nearly all instances. This systemic benefit stems from a fundamental architectural difference: when constraints restrict the search space, greedy algorithms like PICT often make locally optimal choices that lead to dead ends, forcing them to append additional, low-yield test cases just to fulfill outstanding coverage requirements. In contrast, SeqTG natively embeds both inclusionary and exclusionary constraints directly into the decision space of its sequential ILP model. It simultaneously satisfies ``must-include'' requirements while mathematically maximizing the coverage of other pending interactions, effectively turning restrictive constraints into guiding boundaries rather than obstacles.
	
	The absolute dominance of the sequential optimization approach under constrained conditions is conclusively revealed in the robustness metrics. Fig.~\ref{fig:performance_analysis_2}(c) plots the performance profile, showing the cumulative probability of a solver reaching the best-known solution. The curve for SeqTG virtually hugs the y-axis, reaching a probability of nearly 1.0 immediately. This signifies that SeqTG is consistently the absolute best solver for almost every single heavily constrained instance in the dataset. PICT's curve, rising much more slowly, mathematically confirms its tendency to produce suboptimal, bloated solutions when navigating complex constraint landscapes.
	
	This is corroborated by the ranking distribution in Fig.~\ref{fig:performance_analysis_2}(d). In a direct head-to-head comparison across 100 large-scale constrained problems, SeqTG achieved \textbf{Rank 1} (producing the smallest or tied-for-smallest test suite) in an overwhelming \textbf{99\%} of the cases. This near-perfect ranking proves that by mathematically looking ahead and proving the optimal combination for the \textit{next} test case at every step, SeqTG successfully eliminates the late-stage bloat inherent in heuristic methods. This robust constraint-handling capability makes SeqTG particularly indispensable for highly configurable, safety-critical systems where testing resources are strictly limited and validity rules are non-negotiable.
	
	\subsection{Computational Overhead and Warm-Start Efficiency}
	\label{sec:time_evaluation}
	
	Exact ILP solvers inherently introduce computational overhead. To prevent the combinatorial explosion that arises when solving massive models entirely from scratch, we propose a \textit{Warm-Start} strategy: SeqTG retains a fixed fraction $\alpha$ of the PICT-generated test cases to greedily cover ``easy'' interactions, invoking the ILP engine only to optimally cover the residual ones. We evaluate two configurations across 99 industrial benchmarks: $\alpha{=}0.8$ and $\alpha{=}0.9$, retaining 80\% and 90\% of PICT's output, respectively. Table~\ref{tab:seqtg_summary} highlights the time--quality Pareto frontier established by these configurations.
	
	\begin{table}[ht]
		\centering
		\caption{Comparative summary of SeqTG Warm-Start configurations versus the PICT baseline. Times are wall-clock seconds.}
		\label{tab:seqtg_summary}
		\renewcommand{\arraystretch}{1.2}
		\setlength{\tabcolsep}{4pt}
		\begin{tabular}{@{} l c c @{}}
			\toprule
			\textbf{Metric} & $\alpha{=}0.8$& $\alpha{=}0.9$\\
			\midrule
			Warm-start retention       & 80\% of PICT       & 90\% of PICT       \\
			ILP residual to solve      & $\sim$20\%         & $\sim$10\%         \\
			Median generation time     & $\sim$5{,}500\,s   & $\sim$35\,s        \\
			Avg.\ suite reduction vs.\ PICT & $\sim$2.9\%   & $\sim$1.5\%        \\
			\bottomrule
		\end{tabular}
	\end{table}
	
	For the $\alpha{=}0.8$ configuration, the ILP engine must solve the final $\sim$20\% of interactions. Generation times span 500 to 10{,}000\,s, with a median of approximately 5{,}500\,s (roughly 1.5\,hours). A small tail of pathological instances, such as one case requiring $\sim$31{,}000\,s. This configuration delivers compactness, reducing the PICT baseline by an average of 2.9\% across all benchmarks. Conversely, the $\alpha{=}0.9$ configuration leaves only $\sim$10\% of interactions for the solver, which dramatically shrinks the ILP problem. Generation times collapse to 10--500\,s for the vast majority of instances, yielding a median of just $\sim$35\,s---two orders of magnitude faster than $\alpha{=}0.8$. While the average suite size reduction drops slightly to 1.5\%, this extreme speed makes $\alpha{=}0.9$ highly suitable for integration into interactive or continuous-integration workflows.

	\section{Discussion}
	\label{sec:discussion}
	
	The empirical performance of SeqTG underscores the efficacy of its hybrid architecture, which strategically navigates the traditional trade-off between optimality and scalability in combinatorial test generation. By hierarchically decomposing the monolithic NP-hard formulation into smaller, manageable instances of Integer Linear Programming (ILP), our method circumvents the ``combinatorial explosion'' that historically rendered exact optimization infeasible for real-world systems. As demonstrated in our evaluations---spanning the industrial 5G case study to massive constrained benchmarks---SeqTG's optimized local lookahead ensures that each generated test case maximally encompasses new interactions. This approach successfully eradicates the late-stage bloat and myopic outcomes characteristic of classic greedy heuristics, establishing a new state-of-the-art in test suite compactness.
	
	However, despite the success of this sequential decomposition and the computational mitigation provided by our Warm-Start strategy, we must acknowledge the fundamental limitations inherent to exact mathematical programming. Although SeqTG significantly pushes the boundaries of scalability compared to monolithic baselines, exact ILP solvers still face an inherent computational ceiling when applied to ultra-large, highly constrained problem domains. The underlying difficulty stems from the exponential growth of Mixed-Integer Programming (MIP) search trees as the number of parameters, levels, and complex constraints increases. In such extreme scenarios, intricate constraint conflicts severely fragment the feasible solution space, causing traditional branch-and-bound solvers to struggle in finding optimal---or even feasible---solutions within a practical timeframe.
	
	To overcome this ultimate scalability bottleneck, the integration of machine learning (ML), particularly reinforcement learning (RL) \citep{2020Reinforcement,2021Learning}, holds immense promise for the next evolution of combinatorial testing. By learning the underlying structure of complex combinatorial models, RL agents can be trained to dynamically guide the internal decision processes of mathematical solvers. This includes formulating smarter variable-fixing heuristics, optimizing branching strategies, and accelerating constraint propagation along dependency graphs. Hybridizing rigorous mathematical programming with learned, data-driven models represents a highly promising avenue for future research, ensuring the continued practical applicability of exact methods for the next generation of ultra-large-scale software systems.

	\section{Conclusion}
	\label{sec:conclusion}
	
	This paper introduced SeqTG, a novel, scalable, and constraint-aware framework that fundamentally rethinks combinatorial test generation. By strategically decomposing the intractable monolithic ILP formulation into a sequence of tractable, mathematically optimized subproblems, SeqTG successfully bridges the gap between the rapid scalability of greedy heuristics and the rigorous quality of exact solvers. Our evaluations demonstrate that the synergy of the Warm-Start strategy and sequential optimization effectively eradicates the ``late-stage bloat'' that traditionally plagues heuristic algorithms. Across an industrial 5G case study and hundreds of massive, highly constrained configurations, SeqTG consistently generated significantly more compact test suites than leading industry tools such as PICT and IPOG.

	\bibliographystyle{unsrtnat}
    \bibliography{ref}

@book{pressman2014,
  title={Software Engineering: A Practitioner's Approach},
  author={Pressman, R. S. and Maxim, B. R.},
  edition={8th},
  year={2014},
  publisher={McGraw-Hill Education}
}

@article{nie2011,
  title={A survey of combinatorial testing},
  author={Nie, C. and Leung, H.},
  journal={ACM Computing Surveys (CSUR)},
  volume={43},
  number={2},
  pages={1--29},
  year={2011},
  publisher={ACM}
}

@inproceedings{cohen2004,
  title={Constructing test suites for software interaction testing},
  author={Cohen, M. B. and Gibbons, P. B. and Mugridge, W. B. and Colbourn, C. J.},
  booktitle={Proceedings of the 26th International Conference on Software Engineering (ICSE)},
  pages={38--48},
  year={2004},
  organization={IEEE}
}

@article{kuhn2004,
  title={Software fault interactions and implications for software testing},
  author={Kuhn, D. R. and Wallace, D. R. and Gallo, A. M.},
  journal={IEEE Transactions on Software Engineering},
  volume={30},
  number={6},
  pages={418--421},
  year={2004},
  month={Jun},
  publisher={IEEE}
}

@article{1997The,
  title={The AETG system: An approach to testing based on combinatorial design},
  author={Cohen, D. M. and Dalal, S. R. and Fredman, M. L. and Patton, G. C.},
  journal={IEEE Transactions on Software Engineering},
  volume={23},
  number={7},
  pages={437--444},
  year={1997},
  month={Jul},
  publisher={IEEE}
}

@article{Saha2023,
  title={Automated Regression Testing in CI/CD: A Pathway to Higher Reliability and Faster Delivery},
  author={Saha, T. and Chatterjee, S.},
  journal={International Journal of Research Publication and Reviews},
  volume={4},
  number={10},
  pages={2090--2098},
  year={2023},
  month={Oct}
}

@mastersthesis{Otterklau2022,
  title={CI/CD Pipelines with Automated End-to-End Testing},
  author={Otterklau, K.},
  year={2022},
  school={Laurea University of Applied Sciences},
  note={Theseus}
}

@inproceedings{2007IPOG,
  title={IPOG: A general strategy for t-way software testing},
  author={Lei, Y. and Kacker, R. and Kuhn, D. R. and Okun, V. and Lawrence, J.},
  booktitle={Proceedings of the 14th Annual IEEE International Conference and Workshops on the Engineering of Computer-Based Systems (ECBS)},
  pages={549--556},
  year={2007},
  organization={IEEE}
}

@inproceedings{Czerwonka2006,
  title={Pairwise testing in the real world: Practical extensions to test-case scenarios},
  author={Czerwonka, J.},
  booktitle={Proceedings of the 24th Pacific Northwest Software Quality Conference (PNSQC)},
  pages={419--430},
  year={2006}
}

@inproceedings{Shiba2004,
  title={Using artificial life techniques to generate test cases for combinatorial testing},
  author={Shiba, T. and Tsuchiya, T. and Kikuno, T.},
  booktitle={Proceedings of the 28th Annual International Computer Software and Applications Conference (COMPSAC)},
  pages={72--77},
  year={2004},
  organization={IEEE}
}

@article{Ahmed2012,
  title={Application of particle swarm optimization to uniform and variable strength covering array construction},
  author={Ahmed, B. S. and Zamli, K. Z. and Lim, C. P.},
  journal={Applied Soft Computing},
  volume={12},
  number={4},
  pages={1330--1347},
  year={2012},
  publisher={Elsevier}
}

@article{2006Production,
  title={Constraint models for the covering test problem},
  author={Hnich, B. and Prestwich, S. and Selensky, E. and Smith, B. M.},
  journal={Constraints},
  volume={11},
  number={2-3},
  pages={199--219},
  year={2006},
  month={Jul},
  publisher={Springer}
}

@inproceedings{2009IBM,
  title={Model-based testing in practice},
  author={Dalal, S. R. and Jain, A. and Karunanithi, N. and Leitao, J. M. and Lipton, M. and Lott, C. M. and Patton, A. and Horowitz, B. M. and Fredman, M. L.},
  booktitle={Proceedings of the IEEE International Conference on Software Engineering (ICSE)},
  pages={285--294},
  year={2009},
  organization={IEEE}
}

@article{mAETG_ref,
  title={The mAETG strategy for t-way test suite generation},
  author={Zamli, K. Z. and Klaib, M. F. J. and Muthuraman, V. P. M. and others},
  journal={International Journal of Physical Sciences},
  volume={6},
  number={13},
  pages={3175--3188},
  year={2011}
}

@misc{Jenny_ref,
  title={Jenny: A combinatorial test case generator},
  author={Jenkins, B.},
  year={2023},
  note={[Online]. Available: \url{http://burtleburtle.net/bob/math/jenny.html}. [Accessed: Aug. 2023]}
}

@inproceedings{TConfig_ref,
  title={Determination of test configurations for pair-wise interaction coverage},
  author={Williams, A. W.},
  booktitle={Proceedings of the 13th International Conference on Testing Communicating Systems (TestCom)},
  pages={59--74},
  year={2000}
}

@inproceedings{ACA_ref,
  title={Applying ant colony optimization to combinatorial testing},
  author={Chen, X. and Gu, Q. and Qi, J. and Chen, D.},
  booktitle={Proceedings of the 10th International Conference on Quality Software (QSIC)},
  pages={54--63},
  year={2009},
  organization={IEEE}
}

@inproceedings{2020Reinforcement,
  title={Reinforcement learning for integer programming: Learning to cut},
  author={Tang, Y. and Agrawal, S. and Faenza, Y.},
  booktitle={Proceedings of the International Conference on Machine Learning (ICML)},
  year={2020}
}

@misc{2021Learning,
  title={Learning large neighborhood search policy for integer programming},
  author={Wu, Y. and Song, W. and Cao, Z. and Zhang, J.},
  year={2021},
  note={arXiv preprint}
}

\end{document}